\documentclass[11pt,english,a4]{article}

\usepackage{natbib}
\usepackage{amsfonts,amssymb,amsthm,amsmath}
\usepackage[dvips]{graphicx}
\usepackage{syntonly}
\usepackage[activeacute,english]{babel}
\usepackage{setspace,epsfig}
\usepackage{bibentry}
\usepackage[colorlinks=true,linkcolor=blue, citecolor=blue, urlcolor = blue]{hyperref}

\usepackage{booktabs}
\usepackage{multirow}

\begin{document}

\title{\textbf{Treatment effect validation via a permutation test in Stata}\thanks{I wish to thank Dr. Gabriella Conti, Dr Stavros Poupakis and Giacomo Mason from the Department of Economics at UCL for helpful discussions and constructive feedback. Moreover, the author gratefully acknowledge hospitality at the Department of Economics of UCL where this report was prepared. The Stata code for the execution of the permutation test can be found on Github: \textcolor{blue}{https://github.com/masongcm/permtest}. }}
\author{\textbf{Christis Katsouris}\footnote{Ph.D Candidate, Department of Economics, University of Southampton, Southampton, SO17 1BJ, UK. \textit{E-mail Address}: \textcolor{blue}{C.Katsouris@soton.ac.uk}.}  
\\
Department of Economics
\\ 
University of Southampton
\\
\\
Working Paper

}

\date{January, 20 2020}

\maketitle

\begin{center}
\textbf{Abstract} 
\end{center}

In this paper we describe the testing procedure for assessing the statistical significance of treatment effect under the experimental conditions of baseline imbalance across covariates and attrition from the survey, using the permutation tests proposed by \cite{freedman1983nonstochastic} and \cite{romano2016efficient}. We discuss the testing procedure for these hypotheses based on a linear regression model and introduce the new \textit{Stata} command [R] \textbf{permtest} for the implementation of the permutation test in \textit{Stata}. Moreover, we investigate the finite-sample performance as well as the statistical validity of the test with a Monte Carlo simulation study in which we examine the empirical size and power properties under the conditions of baseline imbalance and attrition for a fixed number of permutation steps.


\newpage
\section{Introduction}

Non-parametric methods and in particular permutation tests have been widely used in empirical research as an alternative inference validation methodology especially in finite samples. Permutation tests have been applied to studies of clinical or economic policy interventions which aim to test the treatment effects of a dependent variable e.g., a health outcome. For example, one-sided single-hypothesis block permutation p-values associated with the IPW treatment effect estimate are presented by \cite{campbell2014early} who examine the long-term health effects of the ABC intervention study\footnote{The particular study was designed as a social experiment to investigate whether a stimulating early childhood environment could prevent the development of mild mental retardation in disadvantaged children.}. The small sample size of the study makes the statistical evaluation of the treatment effects challenging. To overcome this problem, the authors use exact (small-sample) permutation tests for the multiple testing of different treatment effects (i.e., in order to exclude spurious treatment effects) using stepdown procedures. 

Furthermore, the appearance of missing data to survey studies due to non-random attrition as well as the possibility of baseline imbalance across the individual characteristics of the participants introduces additional bias in the evaluation of treatment effects. In particular, \cite{kinnamon2014valid} propose the use of exact $\alpha$-level Monte Carlo permutation tests to correct the problem of missing observations in genotype variables, in the absence of other co-founders such as population stratification. The aformentioned experimental conditions clearly affect the statistical inference related to the treatment effect estimate. In this paper, we describe and implement in Stata the permutation test proposed by \cite{freedman1983nonstochastic} and \cite{romano2016efficient} which provides statistical robustification in the inference of the treatment effect estimate especially for the cases of baseline imbalance and attrition. 

\subsection{Related Literature}

Permutation tests are non-parametric randomization tests with superior performance under specific conditions compared with other parametric methods. Specifically, permutation tests are used to adjust the critical values of non-parametric tests (such as the standard t-test for individual statistical significance) and can control for false positives in hypothesis testing even with mild assumptions regarding the underline distribution. The fact that permutation tests solely rely on the randomization of the observed data and not on any underline structure of the prior probability distribution of the error terms, makes them a suitable test for treatment effect validation under baseline imbalance and attrition. 

The underline theory of permutation tests rests on the assumption of an exchangeable error distribution which implies that conditional on its order statistics, the vector of stochastic disturbance terms is uniformly distributed over all its permutations \citep{freedman1983nonstochastic}. In particular, a permutation test is considered to be a conditional test since it generates the permutation distribution conditional on the observed values of the random variable \cite{ernst2004permutation}.  Therefore, a permutation test applied as an alternative to the traditional statistical tests of significance can provide a robust evaluation of the significance of one or more coefficients in multiple regression models. Various studies have shown the use of permutation tests in the statistical validation of treatment effect estimates for clinical trials.

Firstly, in terms of the finite properties and the distributional assumptions of the permutation test, we note here few related papers. In particular, \cite{tanizaki2004small} examine the finite (small sample) properties of the permutation test against the $t$ test for independence between two samples. The author concludes that as expected the t-test based on data generated from normal random variables perform better that the permutation test. However, in the case of fat tailed and skewed distributions the permutation test has much better power performance. Moreover, \cite{solmi2014permutation}, use a permutation test for the evaluation of the treatment effects in alteration single-case experimental designs. In particular, the authors conclude that the permutation test has a robust performance especially in the case of autocorrelation in the error terms of the experimental design since randomization procedures have the statistical flexibility of handling serial dependence in the structure of the error terms (i.e., time varying dependence of the error terms).    

Secondly, the construction of the permutation test varies depending on the experimental conditions under examination. For example, \cite{welch1990construction}, construct a permutation test for block and factorial designs where the authors conclude that the exchangability property of the error terms allows the efficient estimation of the treatment effects in block designs with covariates due to the well mixing property of the distribution of errors between and across blocks. Furthermore, \cite{foster2016permutation}, propose several permutation-based methods which are robust especially in the case of treatment-by-covariate interactions. This is especially a useful permutation test design for statistical validation of treatment effects in the case we allow the interaction between the treatment effect variable and additional covariates in the model. Finally, \cite{winkler2014permutation} provide a general framework of permutation inference for the general linear model with various permutation strategies and complex experimental designs (such as balanced versus unbalanced designs). 

\section{Treatment Effect Linear Regression Model} 

We define the Treatment Effect Linear Regression Model (TELRM), to be the simple linear regression model for an outcome, (e.g., health outcome), $Y$ and the binary (e.g., 1 or 0) explanatory variable $D$ which indicates the participation to a RCT study. For simplicity, we assume that the TELRM has no additional controls, even though the extension in the presence of additional covariates is discussed further below. 
\begin{align}
\label{eq1}
  y_j = \gamma D_j +  \epsilon_j  \ \text{for} \ j =1,...,n \ \
\end{align}

Let $T_n(X_1,...,X_n)$ be a given test statistic with $T_n : \Omega^n \to \mathbb{R}$ based on $\Omega$-valued random variables. For the asymptotic sample case it is known that a central limit theorem holds for $T_n$ under $H_0$ and a consistent variance estimator $V_n$ for its asymptotic variance is available \citep{janssen2003bootstrap}. However, in the case of finite sample theory in order to obtain accurate results for the test statistic of interest we can use a re-sampling method such as the permutation test. More specifically, we consider a sample $X^{*}_1,...,X^{*}_n$ which is drawn given $X_1,...,X_n$  and consider the resampling test statistic and the data dependent critical values given by
\begin{align}
\label{eq2}
 T^{*}_n = T_n (X^{*}_1,...,X^{*}_n) \text{ and } \ c^{*}_n(X_1,...,X_n).  
\end{align}
with a corresponding asymptotic resampling test version of the significance level $\phi_n =1_{(c_n, \infty)}(Tn)$ given by $\phi^{*}_n = g(X_1,...,X_n)$ for $T_n=c^{*}_n(X_1,...,X_n)$.  

Regardless of the test statistic, p-value is a common measure of evidence against the null hypothesis. The proposed permutation test is used within the potential outcome framework in order to estimate permuted p-values for the significance of the treatment effect estimator under the assumption of block randomization, i.e., $Y(\gamma)  \perp \!\!\! \perp   D | \mathbf{X}$. In particular, we are interested for the one-sided p-value of the null hypothesis of no treatment effect which is given by    
\begin{align}
\label{eq3}
 \hat{p} = \hat{Pr}( \mathbf{t}^{*} > t) = \frac{\#( \mathbf{t}^{*r} > t )}{R}
\end{align}

\newpage

Furthermore, we can use the following econometric specification for the validity of the treatment effect estimate via the permutation test for the TELRM with additional covariates.  
\begin{align}
  y_j = \gamma D_j +  \beta^{'}X  +  \epsilon_j  
\end{align}
where $y_j$ is an $(n \times 1)$ vector of health outcomes, $D_j$ an $(n \times 1)$ vector representing a binary treatment effect indicator variable and $X$ an $(n \times p)$ matrix of covariates, where $p < n$. Moreover, the model unknown parameters are $\gamma$ (a scalar) and $\beta^{'}$ a $(p \times 1)$ vector of coefficients. We also assume that $\epsilon_j$ is a sequence of innovation terms following a particular distribution with at least finite first and second moments, that is, $\epsilon_j \sim ( \mu, \sigma^2)$, which is not necessarily normal. In the case where $\beta=0$ we have the model of treatment effects with no covariates. Also, even though there are no restrictions on the parameter space of $\gamma$, in order to preserve the monotonicity of the power function we use a restricted set as the parameter space of the treatment effect estimator, that is, $\gamma \in (0,1)$.  

To shed more light in the procedure required to construct the permutation test, consider for example the methodology of \cite{freedman1983nonstochastic}, which is considered to be a permutation testing procedure of the adjusted p-values under the reduced model and involves the following steps. First, we regress the outcome on the effect of interest as well as the nuisance parameters and we use the parameter $\beta$ in order to compute the statistic of interest $T_0$ (i.e., the statistics under the full model). Then, we regress the outcome against the reduced model that contains only the nuisance parameters.  Computing the set of permuted outcome values and pre-multiplying the reduced model residuals by a permutation matrix we get the adjusted permuted outcomes. We repeat these steps a large number of times (i.e., number of permutations e.g $M=1000$) to build the reference distribution of the permuted $T^{*}$ statistics under the null hypothesis $H_0$. Finally, by counting the number of times $\{ T^{*}_j \geq T_0 \}$ in order to get the p-value of the permutation test. The permutation test is considered to be very computer-intensive and thus there is increasing need to evaluate its performance for small-sample properties.

In this paper, since we consider two specific experimental conditions we use the methodology of linear conditioning proposed by \citep{freedman1983nonstochastic} which allows to accommodate as a correction for baseline imbalance. Firstly, considering the covariate imbalance case if it is not accommodated correctly in the estimation procedure under the assumption of conditional independence of the treatment effect it can invalidate the statistical inference. Thus, to test for such effects we can consider the matrix of covariates $X$ to include a vector of regressors which are characterised by imbalance in terms of its distributional moments across the control and the treatment group of the experimental design. An alternative design which can accommodate the existence of such effects is to assume that that the covariates included in $X$ are found to be imbalanced across the treatment group. Therefore, in order to correct for this model misspecification we can use the linear conditioning methodology which corrects for the imbalance of regressors.  
  
Secondly, we consider the case of attrition within the experimental design occurring due to the introduction of time factor, and can negate the randomization in the initial experimental design (see, e.g., \cite{seaman2013review},  \cite{hausman1979attrition}). Attrition within an experiment can be due to observables as well as unobservable characteristics of the participants within the study. Furthermore, in order to correct for attrition we consider the methodology of Inverse Probability Weighting, (IPW), proposed by \citep{wooldridge2010econometric}. IPW is a statistical methodology which re-weights the subjects and gives more weight to subjects that present observable characteristics that are more similar to the attributed subjects. Various studies had considered the attrition scenario for permutation tests (e.g., see \cite{tang2009applying}). Moreover, if the probability of attrition is correlated with experimental response, then traditional statistical techniques can lead to biased and inconsistent estimates of the  treatment effect. 

\newpage

\section{The permtest command}

\subsection{Syntax}

The $[R]$ \textbf{permtest} command implements the permutation test \cite{freedman1983nonstochastic} and \cite{romano2016efficient}. The syntax of the command is described below.   

\texttt{permtest} \textit{depvar} \textit{indepvars} $[if]$ $[in]$  $[ , \texttt{treat(\textit{string})} \ \texttt{np}(\textit{integer}) \ \texttt{blockvars}(\textit{varlist}) \ \texttt{lcvars}(\textit{varlist})]$.

The two required arguments of the $[R]$ \textbf{permtest} command are $\textit{depvars}$ which is the dependent variable (outcome variable $y$) and \textit{indepvars} which is the binary treatment effect indicator variable, $D$). The command allows for the dependent variable to be also in a vector form of outcomes e.g., $\mathbf{y} = ( y_1, y_2, y_2 )$ especially when considering the outcomes of the same subjects under different experimental conditions. Detailed instructions for users are given below: 

\begin{enumerate}
\item \texttt{permtest} y, \texttt{treat}(D) \texttt{np}(1000)
\item The inputs should be outcome variable (y), treatment variable (D), number of permutations (np). 

\item When block randomized design is allowed use \texttt{blockvars}(\textit{varlist}). 

\end{enumerate}

\subsection{Options}

We now describe the options of the main command. Notice that when these options are left unspecified, the \textbf{permtest} command executes the procedure based on the default settings. 

\begin{itemize}

\item \texttt{treat(string)} is the binary treatment effect indicator variable of the model which is used to evaluate the linear treatment effect. The specification of the particular variable is essential for the \texttt{permtest} command to run. Furthermore, the treatment effect indicator variable should be an integer for example it could have the assigned values $0$ if the patient is part of the control group and $1$ if the patient is part of the treatment group.    

\item \texttt{np(integer)} is the number of permutations to be used by the \texttt{permtest} command to derive the asymptotic distribution and p-values of the t-test for deciding the statistical significance of the treatment effect. The higher the number of permutations, the higher the complexity and execution time of the permutation test.   

\item \texttt{blockvars(varlist)} is the optional input variable list used to specify a block permutation design. The default permutation design is the naive permutation which overrides blocks. Alternatively, if a variable list is specified the permutation test algorithm uses block randomization techniques based on the variable list of subjects' characteristics such as age, gender or other common variable across subjects.

\item \texttt{lcvars(string)} is the optional input variable list used to specify linear conditioning. This includes all variables which can be used as additional controls. The default is no linear conditioning variables, so the permutation test works with no controls as well. 

\item \texttt{sdmethod} is the stepdown method to be used by the \texttt{permtest} command. The default stepdown method is \texttt{(rw16)} which uses the stepdown hypothesis testing methodology of \cite{romano2016efficient}. Alternatively, the option \texttt{(rp)} implements the stepdown methodology of \cite{campbell2014early}. 

\end{itemize}


\newpage

\begin{itemize}

\item \texttt{naive} Whether to perform naive permutation (overrides blockvars).

\item \texttt{reverse} Whether to reverse ALL outcomes for pvalues and TEs.

\item \texttt{robust} Whether to use robust standard errors.

\item \texttt{link} Probit or Logit link function to be used as an option for the IPW. 

\item \texttt{tipwcovars1} Control variables for IPW (applies to all variables if specified alone).

\item \texttt{verbose} Print full progress of routine.

\item \texttt{effsize} Display effects in terms of effect size. 

\item \texttt{savemat} File to save results matrix. 

\end{itemize}

\subsection{Saved Results}

The command \texttt{permtest} generates the following results in \texttt{e()}:

\underline{Scalars}

\begin{itemize}
\item \texttt{e(pval$\_$asym2s)} two-sided asymptotic pvalues
\item \texttt{e(pval$\_$asym1s)} one-sided asymptotic pvalues
\item \texttt{e(pval$\_$perm)} one-sided permutation pvalues
\item \texttt{e(pval$\_$permipw)} one-sided permutation+IPW pvalues
\item \texttt{r(pval$\_$permipw)} one-sided permutation+IPW pvalues
\item \texttt{e(pval$\_$permsd)} one-sided permutation IPW stepdown pvalues
\item \texttt{e(tstat$\_$p)} Matrix of permuted t-stats
\item \texttt{e(tstat$\_$p$\_$ipw)} Matrix of permuted t-stats with IPW
\item \texttt{e(tstat$\_$s)} Matrix of original t-stats
\item \texttt{e(tstat$\_$s$\_$ipw)} Matrix of original t-stats with IPW
\item \texttt{e(res)} Matrix with table of results.
\end{itemize}

\underline{Matrices}

\begin{itemize}
    \item \texttt{e(ipWeights)} Matrix of inverse probability weights.
    \item \texttt{e(Yperm)} Matrix of permuted.
\end{itemize}

\newpage

\subsection{An Illustrative Example}

Consider the following simulated example that illustrates the finite sample performance of the permutation test proposed by \cite{romano2016efficient}. First, we simulate data to construct the Treatment Effect Linear Regression Model (TELR) using the following code: 
\begin{verbatim}
. set obs 25
. gen d = (rnormal(0,1)<0)
. gen e = rnormal(0,1)
. gen y = 0.8*d + e
\end{verbatim}
Then we regress $y$ on $D$, which includes a model intercept and the binary explanatory covariate:
\begin{verbatim}
. reg y d
\end{verbatim}


Next, we generate a (TELRM) with baseline imbalance in the covariates and attrition and again check the statistical significance  of the treatment effect. 

First, for the experimental condition of baseline imbalance we generate the following code: 

\begin{verbatim}
. gen d  = (rnormal(0,1)<0)
. gen X1 = rnormal(0,2)
. gen rand = runiform()	
. gen X2 = (rand>0.7) + 1
. gen e = rnormal(0,1)
. gen y =  0.8*d + 0.20*X1 + 0.05*X2 + e
\end{verbatim}


Second, for the experimental condition of attrition we generate the following code: 

\begin{verbatim}
. gen W = rnormal(0,1)	
// generate outcomes with treatment effects
. gen y = `gamma'*D + 1.5*W + e
// generate attrition
. gen R = (1 + 0.5*D + 1*W + rnormal(0,1) > 0)		
// R=1 means Y is observed
. replace y = . if R==0
. reg y D W 
\end{verbatim}


\textbf{Remark.} Thus, by observing the statistical significance of the $t-$test under these two experimental conditions is likely that the null hypothesis of no treatment effect is falsely not rejected which is the wrong conclusion since factors such as the small sample size as well as the baseline imbalance and sampling units attrition can invalidate the significance of the treatment effect. Therefore, this motivates the construction of the proposed permutation testing procure. Moreover, we can run again the examples described above for a set of simulated data.  

\begin{verbatim}
permtest y, treat(D) np(`np') ipwcovars1(W)
\end{verbatim}

\newpage

\section{Monte Carlo Simulation}

According to \cite{lalonde1986evaluating}, a reasonable estimator of the average treatment effect should be able to closely replicate the experimental estimate of the effect of interest e.g., a health or economic outcome.  Furthermore, a Monte Carlo simulation study can provide us with useful insights regarding the aforementioned conditions especially in cases where theoretical results are limited. According to \cite{advani2018mostly} a carefully designed and empirically motivated Monte Carlo study is capable of informing the empirical researcher of the actual ranking of various estimators when applied to a given problem using a given data set. Thus, the validity of the permutation test is related to its ability to detect a statistical significant treatment effect estimator of the linear econometric model. Therefore, the Monte Carlo results can provide statistical evidence regarding under which experimental conditions is expected that the permutation test robustly identifies the treatment effect of a related intervention policy. 

When designing a Monte Carlo experiment there are important aspects which should be taken into account for robust estimation of the empirical size and power of the testing procedure under consideration. Related aspects include: \textit{(i)} the number of replications, $(B)$, \textit{(ii)} the econometric specification used in the experimental design \textit{(iii)} the error distribution used to simulate the data generating process (DGP). However, in order to control for certain experimental conditions we can keep some aspects of the experimental design fixed (such as the model specification, and the number of replications) while varying other factors (such as the treatmenet effect coefficient, the sample size and the error distribution). In particular, according to \cite{Davidson1995estimation}, the finite distribution of the test statistic of a regression parameter in the Monte Carlo design does not depend on the variance of the error term and thus we can fix the variance of the error terms at some arbitrary level.   

Empirical size and power of the permutation test are calculated as the percentage of times in the samples which were permuted the null of no treatment effect is rejected. We test various experimental designs with different treatment effect parameters in order to find the optimal design which can give the desirable properties. We consider a sequence of local data generating processes which are not nested within the alternative hypothesis and we check the monotonicity property of the power function as well as for any signs of size distortion. The computation on the empirical size and power are carried out with $B=1000$ replications to control the rate of convergence and $r=1000$ (number of permutations) permutated test statistics from the permutation distribution which is a sufficient number to get an accurate estimate of the exact p-value of the permutation test. Let $D_j$ denote the treatment indicator variable, that is, $D_j=1$ if the individual was assigned to treatment and $D_j=0$ otherwise. The empirical size and power of the Monte Carlo simulation study represent the rejection rates of the following null hypothesis against the alternative hypothesis, that is $H_0: \gamma = 0 \ \text{versus} \ H_1: \gamma > 0$ with significance level $\alpha = 0.05$ and $j=1,..,n$ where $n=\{20,40,60\}$. 

The specific framework allow us to test the validity of the permutation test under different experimental designs by varying the value of the hypothesized treatment effect, the sample size, the error distribution (i.e., symmetric vis-a-vis skewed), the nature of the randomization procedure (i.e., via block permutation), baseline imbalance (i.e., via linear conditioning), selective attrition (i.e., via IPW), as well as multiple hypothesis for the true value of the treatment effect (i.e., via stepdown adjustment of p-values). For comparison purposes, we report the p-values of the t-statistic of the linear regression model computed by expression \eqref{eq3} vis-a-vis the simulated p-value that corresponds to the permutation test on the sample statistics. Furthermore, we employ the method of re-sampling from the conditional distribution which ensures that the test statistics remain exact and distribution free with only a small loss of efficiency.

\newpage
\subsection{Baseline Experimental Design}

For the baseline experimental design after extensive trial and error we choose to use as DGP an econometric specification which has no covariates and a specification which includes two covariates. Furthermore, in order to capture different experimental errors we draw the distributional errors from two symmetric distributions and two skewed distributions for each DGP. 

\textbf{Model 1 (Treatment Effect estimation with no covariates)}
\begin{align}
 y_j = \gamma D_j + \epsilon^{(k)}_j, \ \text{where} \ \epsilon^{(k)}_j \sim  F(\epsilon^{(k)}), \ j=\{1,...,n\}. 
\end{align}
$\epsilon^{(1)}_j \sim N(0,1)$, $\epsilon^{(2)}_j \sim S(1)$, $\epsilon^{(3)}_j \sim W(a=1, b=1.5)$ and $\epsilon^{(4)}_j \sim C(a=1,b=0.25)$.   
\

The above model has a simple specification, however we can test the validity of the permutation test in detecting the significance of the treatment effect estimator without conditioning on covariates and thus can give us a good indication of the suitability of the error distribution used in the model. Table \ref{table1}, presents the empirical size and power of the Monte Carlo experiment for the particular DPG and conditions which represents the number of times the estimated p-values are less than the significance level $\alpha=5\%$. 

As we can see from the results on Table \ref{table1}, firstly the empirical size for the permutation test is close to the nominal size and thus there is no indications of empirical size distortions. Moreover, the permutation test performs as good as the OLS t-statistic under the normality assumption which is a first sign of the validation of the test. Furthermore, since we are looking to identify the experimental conditions under which the permutation test performs better than the t-statistic we check also other error distributions. For example, in the case of Student distribution with $n=40$ and $\gamma > 0.5$ the permutation test has slightly higher power that then standard t-test. However performance differences become more noticeable as the sample sizes are decreased and skewed errors are introduced. More specifically, in the case of the Weibull distribution which produces skewed error terms, the improvement in the power function of the permutation test is seen only in the case of the value of the treatment effect close to the boundary of the restricted parameter space of $\gamma$. Finally, in the case of Cauchy error terms even though the power function is at lower levels for both tests, we can observe much higher values for the permutation test, an indication that it has superior performance under the assumption of extremely skewed experimental errors, which is often the case in randomized experiments due to attrition and non-response in follow-up surveys.  

\textbf{Model 2 (Treatment Effect estimation with covariates)}
\begin{align}
 y_j = \gamma D +  0.20*X_1 + 0.05*X_2 +  \epsilon^{(k)}_j, \ \text{where} \ \epsilon^{(k)}_j \sim  F(\epsilon^{(k)}), \ j=\{1,...,n\}.  
\end{align}
$\epsilon^{(1)}_j \sim N(0,1)$, $\epsilon^{(2)}_j \sim S(1)$, $\epsilon^{(3)}_j \sim W(a=1, b=1.5)$ and $\epsilon^{(4)}_j \sim C(a=1,b=0.25)$. 
\

The above specification provides an identification of the treatmenet effect in a model with covariates. The model includes a continuous covariate i.e., $X_1 \sim N(0,2)$ and a discrete binary covariate which resembles an individual continuous characteristic and a binary characteristic (e.g., gender) respectively. According to \cite{nagelkerke2000estimating}, the treatment effect estimation with covariates adjusts treatment received for any confounding effects and thus this technique can identify the true or causal treatment effect. Firstly, the inclusion of covariates in the econometric specification corrects for any size distortions in the case of the parametric t-test especially in the case of the Cauchy distribution and small sample size. Secondly, even though the power function for the permutation test has generally slightly lower levels than in the case of the model with no covariates in the case of small sample size (i.e., $n=20$) increasing the degree of skewness (i.e, using the Cauchy instead of Weibull distribution) shows improvements in the performance of the permutation test similar to the model with no covariates.

\newpage

\begin{table}[htbp]
  \centering
  \caption{Empirical size and power for $H_0: \gamma = 0$ and $H_0: \gamma > 0$ with model 1 for the baseline experimental design}
    \begin{tabular}{|c|c|c|c|c|c|c|c|c|c|c|c|}
    \hline
          &       & \multicolumn{5}{c}{Parametric p-value} & \multicolumn{5}{|c|}{Permutation p-value} \\
    \hline
    \multicolumn{1}{|c|}{$n$} & \multicolumn{1}{|c|}{$\mathbf{\gamma}$} & N     & S     & W     & L     & C     & N     & S     & W     & L     & C \\
    \hline
    \multicolumn{1}{|c|}{\multirow{9}[18]{*}{20}} & \multicolumn{1}{|c|}{0} & 0.053 & 0.038 & 0.051 & 0.051 & 0.033 & 0.052 & 0.041 & 0.051 & 0.049 & 0.050 \\
    \multicolumn{1}{|c|}{} & \multicolumn{1}{|c|}{0.1} & 0.066 & 0.084 & 0.087 & 0.088 & 0.063 & 0.067 & 0.086 & 0.102 & 0.103 & 0.097 \\
    \multicolumn{1}{|c|}{} & \multicolumn{1}{|c|}{0.2} & 0.106 & 0.113 & 0.136 & 0.161 & 0.095 & 0.104 & 0.124 & 0.154 & 0.179 & 0.140 \\
    \multicolumn{1}{|c|}{} & \multicolumn{1}{|c|}{0.3} & 0.165 & 0.202 & 0.228 & 0.249 & 0.148 & 0.163 & 0.211 & 0.245 & 0.270 & 0.217 \\
    \multicolumn{1}{|c|}{} & \multicolumn{1}{|c|}{0.4} & 0.195 & 0.261 & 0.316 & 0.352 & 0.211 & 0.192 & 0.266 & 0.336 & 0.382 & 0.281 \\
    \multicolumn{1}{|c|}{} & \multicolumn{1}{|c|}{0.5} & 0.260 & 0.371 & 0.397 & 0.444 & 0.266 & 0.259 & 0.386 & 0.413 & 0.462 & 0.349 \\
    \multicolumn{1}{|c|}{} & \multicolumn{1}{|c|}{0.6} & 0.337 & 0.420 & 0.473 & 0.568 & 0.326 & 0.342 & 0.427 & 0.490 & 0.581 & 0.406 \\
    \multicolumn{1}{|c|}{} & \multicolumn{1}{|c|}{0.7} & 0.426 & 0.531 & 0.540 & 0.607 & 0.368 & 0.431 & 0.545 & 0.553 & 0.634 & 0.455 \\
    \multicolumn{1}{|c|}{} & \multicolumn{1}{|c|}{0.8} & 0.499 & 0.629 & 0.617 & 0.682 & 0.417 & 0.499 & 0.640 & 0.628 & 0.697 & 0.493 \\
    \hline
    \multicolumn{1}{|c|}{\multirow{9}[18]{*}{40}} & \multicolumn{1}{|c|}{0} & 0.051 & 0.058 & 0.050 & 0.035 & 0.033 & 0.047 & 0.057 & 0.050 & 0.039 & 0.053 \\
    \multicolumn{1}{|c|}{} & \multicolumn{1}{|c|}{0.1} & 0.098 & 0.107 & 0.107 & 0.100 & 0.060 & 0.096 & 0.104 & 0.109 & 0.103 & 0.093 \\
    \multicolumn{1}{|c|}{} & \multicolumn{1}{|c|}{0.2} & 0.146 & 0.194 & 0.174 & 0.215 & 0.099 & 0.143 & 0.197 & 0.180 & 0.219 & 0.133 \\
    \multicolumn{1}{|c|}{} & \multicolumn{1}{|c|}{0.3} & 0.245 & 0.303 & 0.286 & 0.330 & 0.154 & 0.247 & 0.298 & 0.288 & 0.345 & 0.216 \\
    \multicolumn{1}{|c|}{} & \multicolumn{1}{|c|}{0.4} & 0.318 & 0.408 & 0.403 & 0.487 & 0.217 & 0.316 & 0.408 & 0.404 & 0.509 & 0.284 \\
    \multicolumn{1}{|c|}{} & \multicolumn{1}{|c|}{0.5} & 0.475 & 0.543 & 0.539 & 0.610 & 0.273 & 0.468 & 0.540 & 0.535 & 0.630 & 0.355 \\
    \multicolumn{1}{|c|}{} & \multicolumn{1}{|c|}{0.6} & 0.585 & 0.646 & 0.647 & 0.718 & 0.337 & 0.572 & 0.655 & 0.654 & 0.726 & 0.415 \\
    \multicolumn{1}{|c|}{} & \multicolumn{1}{|c|}{0.7} & 0.697 & 0.748 & 0.756 & 0.802 & 0.389 & 0.695 & 0.755 & 0.759 & 0.809 & 0.480 \\
    \multicolumn{1}{|c|}{} & \multicolumn{1}{|c|}{0.8} & 0.774 & 0.831 & 0.822 & 0.851 & 0.433 & 0.776 & 0.830 & 0.822 & 0.860 & 0.522 \\
    \hline
    \multicolumn{1}{|c|}{\multirow{9}[18]{*}{60}} & \multicolumn{1}{|c|}{0} & 0.052 & 0.054 & 0.050 & 0.046 & 0.024 & 0.050 & 0.059 & 0.052 & 0.052 & 0.038 \\
    \multicolumn{1}{|c|}{} & \multicolumn{1}{|c|}{0.1} & 0.105 & 0.102 & 0.123 & 0.132 & 0.055 & 0.108 & 0.105 & 0.121 & 0.139 & 0.081 \\
    \multicolumn{1}{|c|}{} & \multicolumn{1}{|c|}{0.2} & 0.154 & 0.217 & 0.220 & 0.265 & 0.094 & 0.155 & 0.216 & 0.216 & 0.276 & 0.135 \\
    \multicolumn{1}{|c|}{} & \multicolumn{1}{|c|}{0.3} & 0.320 & 0.352 & 0.350 & 0.415 & 0.142 & 0.315 & 0.357 & 0.355 & 0.420 & 0.196 \\
    \multicolumn{1}{|c|}{} & \multicolumn{1}{|c|}{0.4} & 0.463 & 0.542 & 0.498 & 0.584 & 0.196 & 0.457 & 0.541 & 0.497 & 0.594 & 0.261 \\
    \multicolumn{1}{|c|}{} & \multicolumn{1}{|c|}{0.5} & 0.568 & 0.661 & 0.644 & 0.714 & 0.259 & 0.569 & 0.669 & 0.646 & 0.717 & 0.336 \\
    \multicolumn{1}{|c|}{} & \multicolumn{1}{|c|}{0.6} & 0.720 & 0.761 & 0.747 & 0.816 & 0.317 & 0.723 & 0.758 & 0.748 & 0.816 & 0.397 \\
    \multicolumn{1}{|c|}{} & \multicolumn{1}{|c|}{0.7} & 0.843 & 0.878 & 0.824 & 0.870 & 0.370 & 0.843 & 0.876 & 0.820 & 0.872 & 0.456 \\
    \multicolumn{1}{|c|}{} & \multicolumn{1}{|c|}{0.8} & 0.927 & 0.909 & 0.907 & 0.913 & 0.422 & 0.928 & 0.907 & 0.908 & 0.914 & 0.509 \\
    \hline
    \end{tabular}%
  \label{table1}%
\end{table}%

\underline{Notes:} DGP is given by  $y = \gamma D + \epsilon$ and the significance level of $H_0: \gamma = 0$ is $\alpha=0.05$ with $B=1000$ number of replications based on 1000 number of permutation steps. 

\textbf{Remarks}. In order to be able to correctly compare the statistical effect of $\gamma$ across the different distributions and across the different values of $\gamma$ within the restricted set $\gamma \in [0,1)$  we rescale each distribution with respect its population mean and variance in order to keep the same variance for the given econometric specification. In particular, Student and Weibull errors can be considered to be simulating skewed errors in the experimental design. Moreover, Cauchy errors is considered to be a fat tailed distribution while the Log-Normal to be both fat tailed and skewed. Generally, by observing the results of the empirical size and power of Tables 1 we can say that the empirical size seems to be close to the nominal value for all distributions expect in some cases just a little undersized (e.g., alpha = $3.8\%$). Also, the lognormal and Weibull distributed errors give higher power values in comparison to the normal and student errors (with lognormal giving the highest power) and also show more clear improvements between the p-value obtained from OLS vis-a-vis the p-value obtained with the permutation procedure especially for smaller sample sizes (e.g., $n=20$) which is a desirable result for our simulation study. In terms of the Cauchy errors, even though it has lower power values in comparison with the rest, there is more clear difference between the power of the non-parametric test (i.e., permutation test) and the parametric test.

\newpage

\begin{table}[htbp]
  \centering
  \caption{Empirical size and power for $H_0: \gamma = 0$ and $H_0: \gamma > 0$ with model 2 for the baseline experimental design}
    \begin{tabular}{|c|c|c|c|c|c|c|c|c|c|c|c|}
    \hline
          &       & \multicolumn{5}{|c|}{Parametric p-value} & \multicolumn{5}{|c|}{Permutation p-value} \\
    \hline
    \multicolumn{1}{|c|}{$n$} & \multicolumn{1}{c|}{$\mathbf{\gamma}$} & N     & S     & W     & L     & C     & N     & S     & W     & L     & C \\
    \hline
    \multicolumn{1}{|c|}{\multirow{9}[18]{*}{20}} & \multicolumn{1}{|c|}{0} & 0.049 & 0.038 & 0.038 & 0.049 & 0.041 & 0.054 & 0.040 & 0.041 & 0.050 & 0.043 \\
    \multicolumn{1}{|c|}{} & \multicolumn{1}{|c|}{0.1} & 0.080 & 0.073 & 0.072 & 0.084 & 0.072 & 0.072 & 0.075 & 0.075 & 0.096 & 0.074 \\
    \multicolumn{1}{|c|}{} & \multicolumn{1}{|c|}{0.2} & 0.115 & 0.118 & 0.127 & 0.155 & 0.110 & 0.116 & 0.121 & 0.129 & 0.163 & 0.124 \\
    \multicolumn{1}{|c|}{} & \multicolumn{1}{|c|}{0.3} & 0.143 & 0.164 & 0.194 & 0.232 & 0.162 & 0.145 & 0.177 & 0.199 & 0.248 & 0.179 \\
    \multicolumn{1}{|c|}{} & \multicolumn{1}{|c|}{0.4} & 0.190 & 0.236 & 0.251 & 0.322 & 0.209 & 0.188 & 0.244 & 0.260 & 0.331 & 0.232 \\
    \multicolumn{1}{|c|}{} & \multicolumn{1}{|c|}{0.5} & 0.259 & 0.308 & 0.326 & 0.405 & 0.273 & 0.256 & 0.316 & 0.344 & 0.413 & 0.292 \\
    \multicolumn{1}{|c|}{} & \multicolumn{1}{|c|}{0.6} & 0.335 & 0.385 & 0.417 & 0.480 & 0.316 & 0.337 & 0.393 & 0.427 & 0.493 & 0.347 \\
    \multicolumn{1}{|c|}{} & \multicolumn{1}{|c|}{0.7} & 0.413 & 0.467 & 0.509 & 0.557 & 0.374 & 0.406 & 0.473 & 0.505 & 0.573 & 0.409 \\
    \multicolumn{1}{|c|}{} & \multicolumn{1}{|c|}{0.8} & 0.486 & 0.557 & 0.585 & 0.637 & 0.436 & 0.479 & 0.557 & 0.583 & 0.646 & 0.458 \\
    \hline
    \multicolumn{1}{|c|}{\multirow{9}[18]{*}{40}} & \multicolumn{1}{|c|}{0} & 0.043 & 0.058 & 0.050 & 0.042 & 0.034 & 0.044 & 0.059 & 0.055 & 0.047 & 0.041 \\
    \multicolumn{1}{|c|}{} & \multicolumn{1}{|c|}{0.1} & 0.087 & 0.099 & 0.104 & 0.102 & 0.061 & 0.083 & 0.106 & 0.105 & 0.106 & 0.072 \\
    \multicolumn{1}{|c|}{} & \multicolumn{1}{|c|}{0.2} & 0.146 & 0.173 & 0.169 & 0.190 & 0.092 & 0.140 & 0.174 & 0.176 & 0.198 & 0.114 \\
    \multicolumn{1}{|c|}{} & \multicolumn{1}{|c|}{0.3} & 0.207 & 0.273 & 0.287 & 0.311 & 0.144 & 0.199 & 0.272 & 0.289 & 0.324 & 0.170 \\
    \multicolumn{1}{|c|}{} & \multicolumn{1}{|c|}{0.4} & 0.314 & 0.386 & 0.387 & 0.469 & 0.194 & 0.308 & 0.389 & 0.391 & 0.471 & 0.229 \\
    \multicolumn{1}{|c|}{} & \multicolumn{1}{|c|}{0.5} & 0.449 & 0.506 & 0.495 & 0.596 & 0.256 & 0.445 & 0.512 & 0.502 & 0.599 & 0.291 \\
    \multicolumn{1}{|c|}{} & \multicolumn{1}{|c|}{0.6} & 0.592 & 0.640 & 0.606 & 0.694 & 0.317 & 0.587 & 0.642 & 0.608 & 0.694 & 0.352 \\
    \multicolumn{1}{|c|}{} & \multicolumn{1}{|c|}{0.7} & 0.696 & 0.729 & 0.692 & 0.775 & 0.369 & 0.701 & 0.733 & 0.697 & 0.777 & 0.405 \\
    \multicolumn{1}{|c|}{} & \multicolumn{1}{|c|}{0.8} & 0.786 & 0.809 & 0.770 & 0.827 & 0.414 & 0.781 & 0.811 & 0.768 & 0.832 & 0.453 \\
    \hline
    \multicolumn{1}{|c|}{\multirow{9}[18]{*}{60}} & \multicolumn{1}{|c|}{0} & 0.043 & 0.042 & 0.049 & 0.047 & 0.033 & 0.045 & 0.043 & 0.048 & 0.044 & 0.043 \\
    \multicolumn{1}{|c|}{} & \multicolumn{1}{|c|}{0.1} & 0.091 & 0.096 & 0.106 & 0.106 & 0.061 & 0.091 & 0.096 & 0.115 & 0.111 & 0.074 \\
    \multicolumn{1}{|c|}{} & \multicolumn{1}{|c|}{0.2} & 0.176 & 0.213 & 0.202 & 0.222 & 0.109 & 0.177 & 0.212 & 0.206 & 0.229 & 0.131 \\
    \multicolumn{1}{|c|}{} & \multicolumn{1}{|c|}{0.3} & 0.278 & 0.357 & 0.329 & 0.375 & 0.168 & 0.276 & 0.353 & 0.329 & 0.376 & 0.190 \\
    \multicolumn{1}{|c|}{} & \multicolumn{1}{|c|}{0.4} & 0.409 & 0.503 & 0.466 & 0.533 & 0.232 & 0.402 & 0.501 & 0.464 & 0.535 & 0.265 \\
    \multicolumn{1}{|c|}{} & \multicolumn{1}{|c|}{0.5} & 0.562 & 0.654 & 0.613 & 0.659 & 0.283 & 0.561 & 0.655 & 0.615 & 0.660 & 0.318 \\
    \multicolumn{1}{|c|}{} & \multicolumn{1}{|c|}{0.6} & 0.702 & 0.771 & 0.738 & 0.763 & 0.346 & 0.695 & 0.773 & 0.737 & 0.768 & 0.380 \\
    \multicolumn{1}{|c|}{} & \multicolumn{1}{|c|}{0.7} & 0.815 & 0.861 & 0.844 & 0.833 & 0.394 & 0.821 & 0.859 & 0.841 & 0.840 & 0.432 \\
    \multicolumn{1}{|c|}{} & \multicolumn{1}{|c|}{0.8} & 0.893 & 0.908 & 0.892 & 0.895 & 0.449 & 0.892 & 0.908 & 0.894 & 0.902 & 0.489 \\
    \hline
    \end{tabular}%
  \label{table2}%
\end{table}%

\underline{Notes:} DGP is given by  $y = \gamma D +  0.20*X_1 + 0.05*X_2 +  \epsilon$ and the significance level of $H_0: \gamma = 0$ is $\alpha=0.05$ with $B=1000$ number of replications based on 1000 number of permutation steps.   

\textbf{Remarks.} One of the advantages of using a resampling test in that the proper empirical size is guaranteed regardless of the distribution of the error term and thus the choice of the test statistic can be based solely on the power \citep{kennedy1995randomization}. Therefore, this property allows us to compare the parameteric versus the non-parameteric $t-$tests based on their performance under the alternative hypothesis.

\newpage

\subsection{Experimental Design with unbalanced covariates}

The experimental design of unbalance covariates aims to simulate the scenario of having confounding caused by baseline imbalance observed in a particular sample. Thus, conditional on the observed baseline imbalance the estimated treatment effects are biased which result to inflated type I rates \citep{wei2001analysis}. Thus, we aim to check the performance and validity of the permutation test under an experimental design which contains covariates with unbalanced distributional moments. In other words, we want to check whether the permutation test robustly identifies the treatment effect under potential confounding and bias in the original sample design. In order to simulate this scenario, we use two different econometric specifications. The first model consists of using a continuous covariate in the econometric specification and using different means for the treatment and control groups. We also use an auxiliary design to check the robustness of our results with a second DPG which aims to simulate the case of unbalance on the number of people received the treatment, using a Binomial$(n=1,p=0.3)$.  

\textbf{Model 1 (Unbalanced continuous covariate)}
\begin{align}
 y_j = \gamma D +  0.20*X_1 + 0.05*X_2 +  \epsilon_j.  
\end{align}
where $X_1 \sim N(0.25,2)$ if $d=1$ and $X_1 \sim N(-0.25,2)$ if $d=0$ with $\epsilon^{(1)}_j \sim N(0,1)$, $\epsilon^{(2)}_j \sim S(n-1)$, $\epsilon^{(3)}_j \sim W(a=1, b=1.5)$ and $\epsilon^{(4)}_j \sim C(a=1,b=0.25)$. 
\

\textbf{Model 2 (Unbalanced treatment)}
\begin{align}
 y_j = \gamma D +  0.20*X_1 + 0.05*X_2 +  \epsilon_j, \ j=\{1,...,n\}.  
\end{align}
where $D \sim$ Binomial$(1,0.3)$ and $\epsilon^{(1)}_j \sim N(0,1)$, $\epsilon^{(2)}_j \sim S(n-1)$, $\epsilon^{(3)}_j \sim W(a=1, b=1.5)$ and $\epsilon^{(4)}_j \sim C(a=1,b=0.25)$. 
\

Table \ref{table3} and Table \ref{table4} demonstrate the empirical size and power for the unbalance covariates case for the two DGP under consideration. The results are quite similar which verifies the internal validity of the test with econometric specifications which include both covariates and no covariates. Furthermore, in order to correct for the presence of unbalance covariates in the experimental design and to test the validity of the permutation test we use the method of linear conditioning in order to correct for the existence of imbalanced across the treatment covariates. This particular methodology can be used because the randomization of the outcome values implies the independence assumptions between the outcome and the treatment given the model covariates. Since the outcome $Y$ is an unknown function of the treatment, a vector of observed covariates $\mathbf{X}$, and the unobserved error term then we can say that $Y  \perp \!\!\! \perp   D | \mathbf{X}$ and $\mathbf{B}  \perp \!\!\! \perp   D | \mathbf{X}$ and consequently, $Y  \perp \!\!\! \perp   D | ( \mathbf{X} , \mathbf{B})$. 

The case of unbalance covariates in randomized trials has been examined by other authors which point out the importance of correcting for this experimental scenario which can invalidate statistical inference about the treatment effect under consideration. In particular, \cite{permutt1990testing}, examine the effect of unbalanced covariates in case of covariates which are correlated with the outcome under examination and provide evidence that adjusting for the covariate in such cases provides significant improvements for both the empirical size and power of statistical tests. \cite{morgan2012rerandomization}, examine extensively re-sampling techniques such as a permutation test which provide an easy and intuitive way to improve covariate balance in randomized experiments and furthermore improves estimation precision, results in more powerful tests and narrower confidence intervals.  Furthermore, many studies report the importance of adjusting for the baseline values of parameters in experimental designs. For example, \cite{sichieri2014unbalanced} report that unbalanced data at baseline can occur due to underestimated sample sizes and cluster design and can accommodate for up to $35\%$ of the reviewed studies.


\newpage

\begin{table}[h!]
  \centering
  \caption{Empirical size and power for $H_0: \gamma = 0$ and $H_0: \gamma > 0$ with model 1 for unbalanced covariates experimental design}
    \begin{tabular}{|c|c|c|c|c|c|c|c|c|c|c|c|}
    \hline
          &       & \multicolumn{5}{c}{Parametric p-value} & \multicolumn{5}{|c|}{Permutation p-value} \\
    \hline
    \multicolumn{1}{|c|}{$n$} & \multicolumn{1}{|c|}{$\mathbf{\gamma}$} & N     & S     & W     & L     & C     & N     & S     & W     & L     & C \\
    \hline
    \multicolumn{1}{|c|}{\multirow{6}[12]{*}{20}} & \multicolumn{1}{|c|}{0} & 0.046 & 0.046 & 0.041 & 0.049 & 0.046 & 0.047 & 0.045 & 0.045 & 0.054 & 0.051 \\
    \multicolumn{1}{|c|}{} & \multicolumn{1}{|c|}{0.1} & 0.068 & 0.076 & 0.079 & 0.090 & 0.072 & 0.069 & 0.079 & 0.082 & 0.099 & 0.079 \\
    \multicolumn{1}{|c|}{} & \multicolumn{1}{|c|}{0.2} & 0.106 & 0.116 & 0.124 & 0.153 & 0.112 & 0.106 & 0.119 & 0.122 & 0.174 & 0.119 \\
    \multicolumn{1}{|c|}{} & \multicolumn{1}{|c|}{0.3} & 0.138 & 0.160 & 0.182 & 0.226 & 0.160 & 0.139 & 0.169 & 0.195 & 0.260 & 0.171 \\
    \multicolumn{1}{|c|}{} & \multicolumn{1}{|c|}{0.4} & 0.182 & 0.228 & 0.250 & 0.334 & 0.208 & 0.176 & 0.237 & 0.265 & 0.372 & 0.231 \\
    \multicolumn{1}{|c|}{} & \multicolumn{1}{|c|}{0.5} & 0.248 & 0.302 & 0.325 & 0.455 & 0.267 & 0.254 & 0.317 & 0.340 & 0.471 & 0.293 \\
    \hline
    \multicolumn{1}{|c|}{\multirow{6}[12]{*}{40}} & \multicolumn{1}{|c|}{0} & 0.047 & 0.060 & 0.051 & 0.043 & 0.040 & 0.048 & 0.061 & 0.053 & 0.041 & 0.054 \\
    \multicolumn{1}{|c|}{} & \multicolumn{1}{|c|}{0.1} & 0.082 & 0.102 & 0.108 & 0.098 & 0.069 & 0.078 & 0.104 & 0.112 & 0.101 & 0.083 \\
    \multicolumn{1}{|c|}{} & \multicolumn{1}{|c|}{0.2} & 0.135 & 0.179 & 0.177 & 0.226 & 0.099 & 0.132 & 0.176 & 0.180 & 0.230 & 0.118 \\
    \multicolumn{1}{|c|}{} & \multicolumn{1}{|c|}{0.3} & 0.203 & 0.263 & 0.275 & 0.346 & 0.149 & 0.200 & 0.264 & 0.270 & 0.358 & 0.172 \\
    \multicolumn{1}{|c|}{} & \multicolumn{1}{|c|}{0.4} & 0.318 & 0.386 & 0.380 & 0.436 & 0.194 & 0.313 & 0.388 & 0.386 & 0.450 & 0.227 \\
    \multicolumn{1}{|c|}{} & \multicolumn{1}{|c|}{0.5} & 0.443 & 0.506 & 0.493 & 0.593 & 0.250 & 0.439 & 0.508 & 0.500 & 0.611 & 0.289 \\
    \hline
    \multicolumn{1}{|c|}{\multirow{6}[12]{*}{60}} & \multicolumn{1}{|c|}{0} & 0.037 & 0.039 & 0.043 & 0.043 & 0.036 & 0.039 & 0.036 & 0.045 & 0.041 & 0.050 \\
    \multicolumn{1}{|c|}{} & \multicolumn{1}{|c|}{0.1} & 0.090 & 0.098 & 0.107 & 0.106 & 0.068 & 0.091 & 0.103 & 0.112 & 0.114 & 0.089 \\
    \multicolumn{1}{|c|}{} & \multicolumn{1}{|c|}{0.2} & 0.157 & 0.205 & 0.200 & 0.251 & 0.118 & 0.160 & 0.203 & 0.201 & 0.262 & 0.140 \\
    \multicolumn{1}{|c|}{} & \multicolumn{1}{|c|}{0.3} & 0.269 & 0.340 & 0.333 & 0.421 & 0.164 & 0.271 & 0.339 & 0.338 & 0.423 & 0.196 \\
    \multicolumn{1}{|c|}{} & \multicolumn{1}{|c|}{0.4} & 0.414 & 0.507 & 0.468 & 0.586 & 0.215 & 0.406 & 0.505 & 0.471 & 0.591 & 0.259 \\
    \multicolumn{1}{|c|}{} & \multicolumn{1}{|c|}{0.5} & 0.561 & 0.658 & 0.610 & 0.705 & 0.283 & 0.558 & 0.654 & 0.614 & 0.712 & 0.321 \\
    \hline
    \end{tabular}%
  \label{table3}%
\end{table}%

\begin{table}[h!]
  \centering
  \caption{Empirical size and power for $H_0: \gamma = 0$ and $H_0: \gamma > 0$ with model 2 for unbalanced covariates experimental design}
    \begin{tabular}{|c|c|c|c|c|c|c|c|c|c|c|c|}
    \hline
          &       & \multicolumn{5}{|c|}{Parametric p-value} & \multicolumn{5}{|c|}{Permutation p-value} \\
    \hline
    \multicolumn{1}{|c|}{$n$} & \multicolumn{1}{|c|}{$\mathbf{\gamma}$} & N     & S     & W     & L     & C     & N     & S     & W     & L     & C \\
    \hline    
    \multicolumn{1}{|c|}{\multirow{6}[12]{*}{20}} & \multicolumn{1}{|c|}{0} & 0.052 & 0.037 & 0.039 & 0.047 & 0.039 & 0.054 & 0.038 & 0.042 & 0.050 & 0.043 \\
    \multicolumn{1}{|c|}{} & \multicolumn{1}{|c|}{0.1} & 0.071 & 0.069 & 0.074 & 0.088 & 0.070 & 0.068 & 0.074 & 0.077 & 0.090 & 0.071 \\
    \multicolumn{1}{|c|}{} & \multicolumn{1}{|c|}{0.2} & 0.103 & 0.112 & 0.124 & 0.156 & 0.109 & 0.104 & 0.119 & 0.134 & 0.165 & 0.123 \\
    \multicolumn{1}{|c|}{} & \multicolumn{1}{|c|}{0.3} & 0.140 & 0.169 & 0.188 & 0.228 & 0.160 & 0.140 & 0.178 & 0.198 & 0.247 & 0.182 \\
    \multicolumn{1}{|c|}{} & \multicolumn{1}{|c|}{0.4} & 0.192 & 0.234 & 0.252 & 0.317 & 0.210 & 0.186 & 0.247 & 0.263 & 0.328 & 0.230 \\
    \multicolumn{1}{|c|}{} & \multicolumn{1}{|c|}{0.5} & 0.263 & 0.307 & 0.336 & 0.408 & 0.278 & 0.263 & 0.317 & 0.346 & 0.410 & 0.299 \\
    \hline
    \multicolumn{1}{|c|}{\multirow{6}[12]{*}{40}} & \multicolumn{1}{|c|}{0} & 0.044 & 0.056 & 0.052 & 0.041 & 0.035 & 0.049 & 0.059 & 0.054 & 0.046 & 0.041 \\
    \multicolumn{1}{|c|}{} & \multicolumn{1}{|c|}{0.1} & 0.083 & 0.100 & 0.104 & 0.102 & 0.060 & 0.081 & 0.106 & 0.105 & 0.104 & 0.076 \\
    \multicolumn{1}{|c|}{} & \multicolumn{1}{|c|}{0.2} & 0.144 & 0.173 & 0.174 & 0.192 & 0.097 & 0.141 & 0.177 & 0.179 & 0.195 & 0.115 \\
    \multicolumn{1}{|c|}{} & \multicolumn{1}{|c|}{0.3} & 0.209 & 0.272 & 0.279 & 0.309 & 0.147 & 0.205 & 0.271 & 0.289 & 0.329 & 0.166 \\
    \multicolumn{1}{|c|}{} & \multicolumn{1}{|c|}{0.4} & 0.313 & 0.390 & 0.386 & 0.461 & 0.197 & 0.309 & 0.394 & 0.393 & 0.465 & 0.228 \\
    \multicolumn{1}{|c|}{} & \multicolumn{1}{|c|}{0.5} & 0.445 & 0.502 & 0.498 & 0.597 & 0.252 & 0.452 & 0.511 & 0.503 & 0.602 & 0.294 \\
    \hline
    \multicolumn{1}{|c|}{\multirow{6}[12]{*}{60}} & \multicolumn{1}{|c|}{0} & 0.041 & 0.043 & 0.046 & 0.047 & 0.032 & 0.045 & 0.043 & 0.047 & 0.043 & 0.043 \\
    \multicolumn{1}{|c|}{} & \multicolumn{1}{|c|}{0.1} & 0.097 & 0.099 & 0.104 & 0.106 & 0.059 & 0.092 & 0.098 & 0.110 & 0.111 & 0.076 \\
    \multicolumn{1}{|c|}{} & \multicolumn{1}{|c|}{0.2} & 0.171 & 0.213 & 0.201 & 0.218 & 0.110 & 0.173 & 0.211 & 0.206 & 0.226 & 0.130 \\
    \multicolumn{1}{|c|}{} & \multicolumn{1}{|c|}{0.3} & 0.272 & 0.351 & 0.332 & 0.373 & 0.170 & 0.271 & 0.355 & 0.334 & 0.371 & 0.188 \\
    \multicolumn{1}{|c|}{} & \multicolumn{1}{|c|}{0.4} & 0.407 & 0.503 & 0.468 & 0.535 & 0.233 & 0.404 & 0.500 & 0.467 & 0.537 & 0.260 \\
    \multicolumn{1}{|c|}{} & \multicolumn{1}{|c|}{0.5} & 0.559 & 0.662 & 0.614 & 0.651 & 0.284 & 0.553 & 0.664 & 0.616 & 0.657 & 0.323 \\
    \hline
    \end{tabular}%
  \label{table4}%
\end{table}%


\newpage
\subsection{Experimental Design with attrition}

We are particularly interested in examining the effects of attrition in the true identification of the treatment effects via the usage of the permutation test. We expect that the non-parametric test will have better power performance vis-a-vis the parametric test due to its re-sampling properties. Moreover, by considering different distribution errors we are also accounting for the scenario of having different types of experimental errors as well as attrition in the sample under examination. The commonly occurred types of attrition, that is a pre-inclusion attrition as well as attrition within the sample. Clearly, the effect of attrition in experimental designs will affect the internal validity of the results therefore should be taken into consideration when designing experiments. 

Firstly, pre-inclusion attrition occurs when similar subjects are excluded disproportionately from the research sample in a systematic manner. Secondly, post-inclusion attrition, that is, when subjects drop out of a study after they have been included, can severely jeopardize the validity of the study both internally and externally. As a matter of fact the literature has documented that post-inclusion attrition is a threat to the internal validity of the study as well as the causality of the findings since the researcher cannot infer whether attrition occurred because of self-selection of the subjects or due to research intervention. Moreover, internal validity refers to the design’s capability of ensuring that an observed relation between an independent and a dependent variable is not spurious and that alternative explanations for the observed relation have been controlled and can be ruled out (see e.g., \cite{terenzini1982designing}). In terms of attrition within the two groups, the researcher cannot be sure that the samples are unbiased even if the numbers of drop-outs are equally distributed among groups. Thus, attrition in experimental designs can introduce bias due to unrepresentativeness and a diminished likelihood of identifying reliable differences between control and treatment groups. 

Secondly, using a correct specification to incorporate the effect of experimental attrition, gives the probabilities of retention and attrition, which are probit functions. Thus, our primary goal is to correct for the effects of attrition on the model estimates. Thus, we can see that attrition bias is driven by the parametric econometric specification as well as the probability of attrition vis-a-vis the probability of retention. In particular, if attrition is due to individual characteristics, which means that the attrition is not random, then it may be the case that attrition is happening not because of the randomization method but because of the treatment procedure. Thus, using the permutation test in order to test the statistical significance of the treatment effect of the experiment under the conditions of attrition, can only be effective if we test that the detection of treatment effects under the correction of missing data i.e., via IPW, gives reduced bias in comparison to not correcting for the missingness of the random sample.

Extensive examination of the various forms of attrition and the possible corrections under different scenarios (e.g., attrition based on observables vis-a-vis on unobservable) is presented by \cite{huber2012identification}. Following the authors' framework, we assume that the outcome $Y$ is an unknown function of the treatment, a vector of observed covariates $X$, and an unobserved term U, that is, $Y=\phi(D,X,U)$. In particular, the experimental design has external validity of there is identification of ATE on the entire population despite of existence of attrition and moreover there is internal validity if the ATE based on the respondents is identifiable. For the simulation study of the experimental design with attrition we follow the experimental design described by \cite{huber2012identification}. The authors examine both cases of attrition on observables and attrition based on unobservable. For this simulation study and for the purpose of this paper, we focus on the case of attrition based on observables. This is a common framework to examine causal effects and helps the estimation and identification of the statistical significance of the treatment effects especially in the presence of attrition.

\newpage

Therefore we use the following econometric specification which simulates the existence of attrition in the selected sample. 
\begin{align}
y = \gamma D + \beta W + v 
\end{align} 
and in order to introduce the missingness in the data we use the following formulation
\begin{align}
R_i = 1 \{ \alpha_0 + \alpha_1 D + \alpha_2 W + v > 0 \}
\end{align}

with $v \sim N(0,1)$ and $\alpha_0 = 0.1 , \alpha_1 =0.25, \alpha_2 =3.75$. Thus, according to the particular specification $y_i$ is observed when the binary variable $R_i=1$. However since the variable which drives the attrition, i.e., W is introduced also in the attrition equation then we assume that we work within the framework of selection on observables we can assume further that the error terms of both linear equations are independent and normally distributed. 

In terms of the assumptions for the correct specification and identification of the treatment effects we have that since $u \sim N(0,1)$ and $v \sim N(0,1)$ due to the assumption that attrition is related to observables then we also assume that the two error terms are uncorrelated. Furthermore, in order to ensure external validity of the experimental design and thus identification of the ATE on the entire population, then we further assume that 
\begin{align}
 \Delta_{R=1} = ( E[ Y^1 | R=1 ] - E[ Y^0 | R=1 ]   ) 
\end{align}

Additionally, our experimental design in this section is applicable in finite sample cases as well as large sample sizes. For example, \cite{huang2018identification} study the effect of maternal smoking during pregnancy on birth weight, which often is considered to have finite sample size. In particular, the authors in the specific study use a triangular system with multiple endogenous variables with a binary instrument $Z$. The correct identification of the instrument and the IV regression requires to identify an exogenous source of variation. This framework can be easily extended to include the case of attrition in the econometric specification.

\textbf{Simulation Results}. Table \ref{table5} and Table \ref{table6} demonstrate the empirical size and power of the permutation and normal-t test respectively for the experimental design with attrition using the econometric specification with only the treatment effect parameter. As we can observe, under the assumption of attrition in the experimental design and with no correction for IPW the empirical size appears to be overestimated while the power of the test is underestimated especially for values of $\gamma$ between 0.1 and 0.5. 

Furthermore, when comparing the permutation test with the non-parametric t-test under the experimental conditions of attrition we observe a similar performance for empirical size and power of the test, however when we focus on the bias after correcting for attrition with the IPW method, we confirm that the IPW methodology gives treatment estimates with the desirable distributional properties. Finally we compare the bias and MSE for the attrition scenario to observe whether indeed there is a reduction in the bias of the estimation of the treatment effect with the permutation test for finite (small) sample sizes. As we can see from the results of the permutation test, the test under the conditions of attrition works well when we correct for the presence of attrition in the sample with the use of Inverse probability weighting and especially under the assumption that the error distribution follow a Weibull distribution (i.e., under the existence of skewed sampling data). Furthermore, with our simulation study we have show the exaggerated effect of inflated type I error of the method proposed by \cite{romano2016efficient}  under the experimental scenario of attrition and have shown the validity of their methodology under the baseline experimental design as well as the unbalanced covariates design. 

\newpage

\begin{table}[h!]
  \centering
  \caption{Empirical size and power for $H_0: \gamma = 0$ and $H_0: \gamma > 0$ of permutation test for experimental design with attrition}
    \begin{tabular}{|c|c|c|c|c|c|c|c|c|c|c|c|}
    \hline
          & \multicolumn{1}{|c|}{} & \multicolumn{10}{|c|}{Permutation test} \\
    \hline
          &       & \multicolumn{2}{|c|}{N} & \multicolumn{2}{|c|}{S} & \multicolumn{2}{|c|}{W} & \multicolumn{2}{|c|}{L} & \multicolumn{2}{|c|}{C} \\
    \hline
    \multicolumn{1}{|c|}{$n$} & \multicolumn{1}{|c|}{$\mathbf{\gamma}$} & no ipw & ipw   & no ipw & ipw   & no ipw & ipw   & no ipw & ipw   & no ipw & ipw \\
    \hline   
     \multicolumn{1}{|c|}{\multirow{8}[12]{*}{20}} & \multicolumn{1}{|c|}{0} & 0.051 & 0.019 &	0.060 &	0.016 &	0.045  &	0.020 &	0.055 &	0.018 &	0.054 &	0.021 \\
    \multicolumn{1}{|c|}{} & \multicolumn{1}{|c|}{0.1} & 0.060	& 0.025 & 0.069 &	0.023 &	0.058 &	0.026 &	0.065 & 0.023 & 0.057 &	0.026 \\
    \multicolumn{1}{|c|}{} & \multicolumn{1}{|c|}{0.2} & 0.073	& 0.031 & 0.079 &   0.027 &	0.065 &	0.029 &	0.071 &	0.025 &	0.068 &	0.033  \\
    \multicolumn{1}{|c|}{} & \multicolumn{1}{|c|}{0.3} & 0.082	& 0.036	& 0.086	& 0.034	& 0.075	& 0.034	& 0.087	& 0.032	& 0.075	& 0.040 \\
    \multicolumn{1}{|c|}{} & \multicolumn{1}{|c|}{0.4} & 0.098 &	0.050 &	0.097 &	0.044 &	0.089 &	0.047 &	0.099 &	0.040	& 0.089 & 	0.047  \\
    \multicolumn{1}{|c|}{} & \multicolumn{1}{|c|}{0.5} & 0.110	& 0.061 &	0.106 &	0.051 &	0.104 &	0.051 &	0.110 &	0.044	& 0.095 &	0.056 \\
     \multicolumn{1}{|c|}{} & \multicolumn{1}{|c|}{0.6} & 0.121 & 0.072	& 0.121 & 0.063 &	0.124 &	0.065 &	0.127 &	0.052	& 0.107	& 0.065 \\
      \multicolumn{1}{|c|}{} & \multicolumn{1}{|c|}{0.7} & 0.137 &	0.080 &	0.142 &	0.074 &	0.144 &	0.076 &	0.141 &	0.069	& 0.123	& 0.074 \\
       \multicolumn{1}{|c|}{} & \multicolumn{1}{|c|}{0.8} & 0.155 &	0.095 &	0.160 &	0.089 &	0.158 &	0.091 &	0.157 &	0.087	& 0.140 & 0.084 \\
    \hline
    \multicolumn{1}{|c|}{\multirow{8}[12]{*}{40}} & \multicolumn{1}{|c|}{0} & 0.079 & 0.030 & 0.076 & 0.028 & 0.076 & 0.034 & 0.079 & 0.030 & 0.080 & 0.032 \\
    \multicolumn{1}{|c|}{} & \multicolumn{1}{|c|}{0.1} & 0.088 & 0.040 & 0.090 & 0.035 & 0.099 & 0.052 & 0.088 & 0.040 & 0.102 & 0.041 \\
    \multicolumn{1}{|c|}{} & \multicolumn{1}{|c|}{0.2} & 0.100 & 0.047 & 0.107 & 0.042 & 0.117 & 0.060 & 0.100 & 0.047 & 0.111 & 0.052 \\
    \multicolumn{1}{|c|}{} & \multicolumn{1}{|c|}{0.3} & 0.120 & 0.062 & 0.126 & 0.061 & 0.141 & 0.077 & 0.137 & 0.058 & 0.133 & 0.072 \\
    \multicolumn{1}{|c|}{} & \multicolumn{1}{|c|}{0.4} & 0.142 & 0.075 & 0.144 & 0.082 & 0.159 & 0.092 & 0.162 & 0.078 & 0.155 & 0.089 \\
    \multicolumn{1}{|c|}{} & \multicolumn{1}{|c|}{0.5} & 0.174 & 0.092 & 0.167 & 0.103 & 0.188 & 0.110 & 0.188 & 0.095 & 0.179 & 0.110 \\
    \multicolumn{1}{|c|}{} & \multicolumn{1}{|c|}{0.6} & 0.194 & 0.120 & 0.199 & 0.126 & 0.233 & 0.139 & 0.209 & 0.120 &	0.192 &	0.136 \\
      \multicolumn{1}{|c|}{} & \multicolumn{1}{|c|}{0.7} & 0.220 & 0.150 &	0.234 &	0.170 &	0.256 &	0.172 &	0.236 &	0.144	& 0.220	& 0.161 \\
       \multicolumn{1}{|c|}{} & \multicolumn{1}{|c|}{0.8} & 0.250 &	0.180 &	0.272 &	0.213 &	0.292 &	0.215 &	0.267 &	0.184	& 0.244 &	0.192 \\
    \hline
    \multicolumn{1}{|c|}{\multirow{8}[12]{*}{60}} & \multicolumn{1}{|c|}{0} & 0.084 & 0.025 & 0.090 & 0.026 & 0.099 & 0.024 & 0.085 & 0.015 & 0.081 & 0.035 \\
    \multicolumn{1}{|c|}{} & \multicolumn{1}{|c|}{0.1} & 0.102 & 0.036 & 0.109 & 0.037 & 0.119 & 0.035 & 0.111 & 0.028 & 0.096 & 0.044 \\
    \multicolumn{1}{|c|}{} & \multicolumn{1}{|c|}{0.2} & 0.135 & 0.061 & 0.130 & 0.054 & 0.143 & 0.053 & 0.133 & 0.043 & 0.120 & 0.057 \\
    \multicolumn{1}{|c|}{} & \multicolumn{1}{|c|}{0.3} & 0.161 & 0.086 & 0.159 & 0.075 & 0.181 & 0.079 & 0.162 & 0.067 & 0.141 & 0.071 \\
    \multicolumn{1}{|c|}{} & \multicolumn{1}{|c|}{0.4} & 0.203 & 0.113 & 0.185 & 0.102 & 0.210 & 0.115 & 0.194 & 0.102 & 0.162 & 0.101 \\
    \multicolumn{1}{|c|}{} & \multicolumn{1}{|c|}{0.5} & 0.233 & 0.156 & 0.225 & 0.141 & 0.245 & 0.162 & 0.242 & 0.135 & 0.196 & 0.130 \\
    \multicolumn{1}{|c|}{} & \multicolumn{1}{|c|}{0.6} & 0.282 & 0.193 & 0.256 & 0.185 & 0.287 & 0.202 & 0.285 & 0.197 &	0.229 & 0.167 \\
      \multicolumn{1}{|c|}{} & \multicolumn{1}{|c|}{0.7} & 0.328 &	0.241 &	0.309 &	0.232 &	0.330 &	0.245 &	0.338 &	0.256	& 0.265 & 0.204 \\
       \multicolumn{1}{|c|}{} & \multicolumn{1}{|c|}{0.8} & 0.369 &	0.296 &	0.343 &	0.289 &	0.377 &	0.303 &	0.387 &	0.325	& 0.308 & 0.249 \\
    \hline
    \end{tabular}%
  \label{table5}%
\end{table}%

\textbf{Remarks.} To correctly use attrition weighting for a permutation test, a good strategy is to investigate its suitability based on alternative methods, such as to permutate the treatment indicator variable with conditioning on the covariates vis-a-vis without conditioning. Specifically, \cite{tang2009applying} propose to first permutes the treatment status based on the original randomization design, and then reconstruct attrition weights using the permutated data. Moreover, the particular methodology re-runs weighted regressions using the reconstructed weights for increased accuracy. A commonly used practise in the literature is to test for attrition bias by comparing for differences in the estimates of the model via simulation experiments, that is, the performance of the test as seen via the empirical size results, when a correction for attrition is made under the null hypothesis of no attrition.

\newpage
\begin{table}[h!]
  \centering
  \caption{Empirical size and power for $H_0: \gamma = 0$ and $H_0: \gamma > 0$ of OLS test for experimental design with attrition}
    \begin{tabular}{|c|c|c|c|c|c|c|c|c|c|c|c|}
    \hline
          & \multicolumn{1}{|c|}{} & \multicolumn{10}{|c|}{Parametric (OLS) Test} \\
    \hline
          &       & \multicolumn{2}{|c|}{N} & \multicolumn{2}{|c|}{S} & \multicolumn{2}{|c|}{W} & \multicolumn{2}{|c|}{L} & \multicolumn{2}{|c|}{C} \\
    \hline      
    \multicolumn{1}{|c|}{$n$} & \multicolumn{1}{|c|}{$\mathbf{\gamma}$} & no ipw & ipw   & no ipw & ipw   & no ipw & ipw   & no ipw & ipw   & no ipw & ipw \\
        \hline
    \multicolumn{1}{|c|}{\multirow{8}[12]{*}{20}} & \multicolumn{1}{|c|}{0} & 0.052 &	0.053 &	0.064 &	0.048 &	0.050 &	0.049	& 0.058	& 0.044	& 0.045	& 0.041 \\
    \multicolumn{1}{|c|}{} & \multicolumn{1}{|c|}{0.1} & 0.063 & 0.066 & 0.070 & 0.057 & 0.061 & 0.063 & 0.064 & 0.051	& 0.051 &	0.050 \\
    \multicolumn{1}{|c|}{} & \multicolumn{1}{|c|}{0.2} & 0.074 & 0.077 & 0.080 & 0.066 & 0.072 & 0.073 & 0.078 & 0.059 &	0.058 &	0.057 \\
    \multicolumn{1}{|c|}{} & \multicolumn{1}{|c|}{0.3} & 0.084 & 0.089 & 0.084 & 0.072 & 0.077 & 0.085 & 0.088 & 0.074 &	0.076 &	0.067 \\
    \multicolumn{1}{|c|}{} & \multicolumn{1}{|c|}{0.4} & 0.099 & 0.097 & 0.093 & 0.084 & 0.090 & 0.095 & 0.098 & 0.083 &	0.090 &	0.078 \\
    \multicolumn{1}{|c|}{} & \multicolumn{1}{|c|}{0.5} & 0.113 & 0.109 & 0.109 & 0.106 & 0.110 & 0.114 & 0.109 & 0.104 &	0.097 &	0.088 \\
     \multicolumn{1}{|c|}{} & \multicolumn{1}{|c|}{0.6} & 0.121 & 0.119 & 0.122 & 0.124 & 0.122 & 0.136 & 0.124 & 0.117	& 0.111 &	0.104 \\
      \multicolumn{1}{|c|}{} & \multicolumn{1}{|c|}{0.7} & 0.141 &	0.138 &	0.143 &	0.140 &	0.143 &	0.157 &	0.140 &	0.133	& 0.126 &	0.118 \\
       \multicolumn{1}{|c|}{} & \multicolumn{1}{|c|}{0.8} & 0.155 &	0.154 &	0.164 &	0.157 &	0.161 &	0.181 &	0.157 &	0.148 &	0.141 &	0.131  \\
    \hline
    \multicolumn{1}{|c|}{\multirow{8}[12]{*}{40}} & \multicolumn{1}{|c|}{0} & 0.075 & 0.036 & 0.079 & 0.030 & 0.074 & 0.048 & 0.075 & 0.036 & 0.071 & 0.033 \\
    \multicolumn{1}{|c|}{} & \multicolumn{1}{|c|}{0.1} & 0.087 & 0.041 & 0.089 & 0.040 & 0.099 & 0.056 & 0.087 & 0.041 & 0.084 & 0.042 \\
    \multicolumn{1}{|c|}{} & \multicolumn{1}{|c|}{0.2} & 0.109 & 0.059 & 0.104 & 0.061 & 0.122 & 0.074 & 0.109 & 0.059 & 0.109 & 0.053 \\
    \multicolumn{1}{|c|}{} & \multicolumn{1}{|c|}{0.3} & 0.120 & 0.075 & 0.125 & 0.079 & 0.136 & 0.083 & 0.132 & 0.071 & 0.125 & 0.070 \\
    \multicolumn{1}{|c|}{} & \multicolumn{1}{|c|}{0.4} & 0.141 & 0.094 & 0.142 & 0.099 & 0.158 & 0.102 & 0.160 & 0.093 & 0.142 & 0.092 \\
    \multicolumn{1}{|c|}{} & \multicolumn{1}{|c|}{0.5} & 0.172 & 0.109 & 0.166 & 0.124 & 0.186 & 0.136 & 0.187 & 0.144 & 0.165 & 0.114 \\
     \multicolumn{1}{|c|}{} & \multicolumn{1}{|c|}{0.6} & 0.195 & 0.136 & 0.199 & 0.156 & 0.225	& 0.168 & 0.204 & 0.137 &	0.183 &	0.144 \\
      \multicolumn{1}{|c|}{} & \multicolumn{1}{|c|}{0.7} & 0.222 &	0.172 &	0.231 &	0.199 &	0.263 &	0.204 &	0.235 &	0.169	& 0.199 & 0.173 \\
       \multicolumn{1}{|c|}{} & \multicolumn{1}{|c|}{0.8} & 0.245 &	0.205 &	0.265 &	0.243 &	0.297 &	0.245 &	0.266 &	0.210	& 0.218 & 0.201 \\
    \hline
    \multicolumn{1}{|c|}{\multirow{8}[12]{*}{60}} & \multicolumn{1}{|c|}{0} & 0.083 & 0.029 & 0.088 & 0.026 & 0.097 & 0.028 & 0.085 & 0.019 & 0.064 & 0.025 \\
    \multicolumn{1}{|c|}{} & \multicolumn{1}{|c|}{0.1} & 0.104 & 0.052 & 0.107 & 0.045 & 0.119 & 0.042 & 0.108 & 0.032 & 0.084 & 0.038 \\
    \multicolumn{1}{|c|}{} & \multicolumn{1}{|c|}{0.2} & 0.135 & 0.068 & 0.126 & 0.063 & 0.147 & 0.062 & 0.131 & 0.052 & 0.097 & 0.052 \\
    \multicolumn{1}{|c|}{} & \multicolumn{1}{|c|}{0.3} & 0.166 & 0.093 & 0.159 & 0.083 & 0.181 & 0.079 & 0.167 & 0.078 & 0.121 & 0.068 \\
    \multicolumn{1}{|c|}{} & \multicolumn{1}{|c|}{0.4} & 0.203 & 0.127 & 0.185 & 0.113 & 0.209 & 0.128 & 0.196 & 0.118 & 0.138 & 0.091 \\
    \multicolumn{1}{|c|}{} & \multicolumn{1}{|c|}{0.5} & 0.237 & 0.162 & 0.216 & 0.152 & 0.246 & 0.178 & 0.242 & 0.156 & 0.176 & 0.124 \\
     \multicolumn{1}{|c|}{} & \multicolumn{1}{|c|}{0.6} & 0.280	& 0.208	& 0.249 & 0.201 & 0.285 & 0.217	& 0.283 & 0.213 &	0.200 &	0.159 \\
      \multicolumn{1}{|c|}{} & \multicolumn{1}{|c|}{0.7} & 0.326 &	0.259 &	0.306 &	0.249 &	0.333 &	0.259 &	0.336 &	0.272	& 0.234 & 0.192 \\
       \multicolumn{1}{|c|}{} & \multicolumn{1}{|c|}{0.8} & 0.366 &	0.310 &	0.346 &	0.309 &	0.387 &	0.321 &	0.384 &	0.339 &	0.272 & 0.238 \\
    \hline
    \end{tabular}%
  \label{table6}%
\end{table}%

\underline{Notes:} DGP is given by $y = \gamma D + \epsilon$ and the significance level of $H_0: \gamma = 0$ is $\alpha=0.05$ with $B=1000$ the number of replications based on 1000 number of permutation steps. 

\medskip

\textbf{Remarks.} Notice that when comparing the permutation test with the non-parametric $t-$test under the experimental conditions of attrition, we observe a similar performance for empirical size and power for the $t-$test. However when we focus on the bias after correcting for attrition with the IPW method, we confirm that the IPW methodology gives treatment estimates with the desirable distributional properties. Notice that attrition which is related only to the exogenous variables in a structural model does not lead to biased estimates, since these variables are controlled for in the statistical analysis. However when attrition it is driven by endogenous variables then biased estimates can occur. In particular, \cite{doyle2017first}  examine more extensively the effect of attrition bias in the estimation of the treatment effects by using the IPW methodology to adjust for differential attrition and non-response in the survey study. The authors conclude that since there is minimal evidence of contamination across the high and low treatment groups then the internal validity of the test remains quite high.

\newpage

\section{Conclusion}

In this section we discuss main concluding remarks of the Monte Carlo simulation study in relation to the performance of the permutation test under various experimental designs and conditions. In summary, when we compare the permutation test with the parametric $t-$test we observe an improved power performance for the permutation test in the the case of fat tailed and skewed distributions in contrast to the parametric $t-$test. Under the experimental conditions of attrition the IPW procedure reduces the bias of the treatment effect estimator. 

More precisely, the randomization with the use of the permutation test balances out the covariates between the treatment group and the control group due to the law of large numbers. Furthermore, the error distributions we employ simulate different experimental scenarios under which allows us to examine the validity and testing performance of the permutation test in correctly detecting the treatment effects based on the chosen econometric specification.  Our Monte Carlo simulation study which consists of the construction of Freedman-Lane p-values follow similar assumptions and procedure as \cite{heckman2010analyzing} where the authors use small-sample permutation and estimate familywise error rates as well as accounting for compromises in the randomization protocol by conditioning
on background variables to control for the violations of the initial randomization protocol and imbalanced background variables. 

Firstly, as we can observe from the results of the empirical size and power for both tests and especially for the permutation test there is adequate control over Type I error without particular dependence on the underline error distribution. This is a particular useful result in the cases of finite samples with fat tail distributions. Also, the permutation test performs as good as the $t$ test even under the Gaussian error term assumption. More importantly though, similar to \cite{tanizaki2004small}, we conclude that the permutation test has better performance than the $t-$test when the sampling distribution of the error term of the model is not Gaussian especially in the case of interest which is the finite sample case. Specifically, the permutation test is both reliable under the null and powerful under the alternative especially in the case of Cauchy errors (due to its capability of capturing asymmetric effects) as well as in the case of attrition and baseline imbalance. Therefore, we conclude that the permutation test would be a robust finite-sample testing procedure for the evaluation of studies such as the RADIEL.

Secondly, the problem of attrition bias may introduced incosistent model estimates. Thus, correcting these bias effects using the IPW methodology allow us to obtain robust conclusions regarding the population represented in the original sample based on the observed follow-up sample. The nature of attrition affects the internal and external validity of the of the treatment effects. Therefore in experiments of medical intervention or economic policy is crucial for the researcher to understand the possible reasons of attrition of participants as the methodology employed to correct for the presence of attrition in the econometric model will affect the accuracy of the results. For instance, when attrition is related to unobservable factors then a correction for attrition would require an instrument which can capture the unobservable variables which contribute to this experimental condition (e.g., age or some other characteristic which do not affect the decision to take the treatment but drives the attrition from the study). Nevertheless, our simulation experiments suggest that the permuation procedure can provide robust estimates for the conditions under examination in this paper.

\bigskip

\textbf{Acknowledgements}. The author gratefully acknowledge financial support from ERC research grant DynaHEALTH offered to Associate Professor Gabriella Conti at the Department of Economics of UCL. The author declares no conflicts of interests.

\newpage

\newpage

\bibliographystyle{apalike}
\bibliography{myreferences}

\begin{thebibliography}{}

\bibitem[Advani et~al., 2018]{advani2018mostly}
Advani, A., Kitagawa, T., and S{\l}oczy{\'n}ski, T. (2018).
\newblock Mostly harmless simulations? on the internal validity of empirical
  monte carlo studies.
\newblock {\em arXiv preprint arXiv:1809.09527}.

\bibitem[Campbell et~al., 2014]{campbell2014early}
Campbell, F., Conti, G., Heckman, J.~J., Moon, S.~H., Pinto, R., Pungello, E.,
  and Pan, Y. (2014).
\newblock Early childhood investments substantially boost adult health.
\newblock {\em Science}, 343(6178):1478--1485.

\bibitem[Doyle, 2017]{doyle2017first}
Doyle, O. (2017).
\newblock The first 2,000 days and child skills: Evidence from a randomized
  experiment of home visiting.
\newblock Technical report, Working Paper Series.

\bibitem[Ernst et~al., 2004]{ernst2004permutation}
Ernst, M.~D. et~al. (2004).
\newblock Permutation methods: a basis for exact inference.
\newblock {\em Statistical Science}, 19(4):676--685.

\bibitem[Foster et~al., 2016]{foster2016permutation}
Foster, J.~C., Nan, B., Shen, L., Kaciroti, N., and Taylor, J.~M. (2016).
\newblock Permutation testing for treatment--covariate interactions and
  subgroup identification.
\newblock {\em Statistics in biosciences}, 8(1):77--98.

\bibitem[Freedman and Lane, 1983]{freedman1983nonstochastic}
Freedman, D. and Lane, D. (1983).
\newblock A nonstochastic interpretation of reported significance levels.
\newblock {\em Journal of Business \& Economic Statistics}, 1(4):292--298.

\bibitem[Hausman and Wise, 1979]{hausman1979attrition}
Hausman, J.~A. and Wise, D.~A. (1979).
\newblock Attrition bias in experimental and panel data: the gary income
  maintenance experiment.
\newblock {\em Econometrica: Journal of the Econometric Society}, pages
  455--473.

\bibitem[Heckman et~al., 2010]{heckman2010analyzing}
Heckman, J., Moon, S.~H., Pinto, R., Savelyev, P., and Yavitz, A. (2010).
\newblock Analyzing social experiments as implemented: A reexamination of the
  evidence from the highscope perry preschool program.
\newblock {\em Quantitative economics}, 1(1):1--46.

\bibitem[Huang et~al., 2018]{huang2018identification}
Huang, L., Khalil, U., and Y{\i}ld{\i}z, N. (2018).
\newblock Identification and estimation of a triangular model with multiple
  endogenous variables and insufficiently many instrumental variables.
\newblock {\em Journal of Econometrics}.

\bibitem[Huber, 2012]{huber2012identification}
Huber, M. (2012).
\newblock Identification of average treatment effects in social experiments
  under alternative forms of attrition.
\newblock {\em Journal of Educational and Behavioral Statistics},
  37(3):443--474.

\bibitem[Janssen et~al., 2003]{janssen2003bootstrap}
Janssen, A., Pauls, T., et~al. (2003).
\newblock How do bootstrap and permutation tests work?
\newblock {\em The Annals of statistics}, 31(3):768--806.

\bibitem[Kennedy, 1995]{kennedy1995randomization}
Kennedy, F.~E. (1995).
\newblock Randomization tests in econometrics.
\newblock {\em Journal of Business \& Economic Statistics}, 13(1):85--94.

\bibitem[Kinnamon and Martin, 2014]{kinnamon2014valid}
Kinnamon, D.~D. and Martin, E.~R. (2014).
\newblock Valid monte carlo permutation tests for genetic case-control studies
  with missing genotypes.
\newblock {\em Genetic epidemiology}, 38(4):325--344.

\bibitem[LaLonde, 1986]{lalonde1986evaluating}
LaLonde, R.~J. (1986).
\newblock Evaluating the econometric evaluations of training programs with
  experimental data.
\newblock {\em The American economic review}, pages 604--620.

\bibitem[Morgan et~al., 2012]{morgan2012rerandomization}
Morgan, K.~L., Rubin, D.~B., et~al. (2012).
\newblock Rerandomization to improve covariate balance in experiments.
\newblock {\em The Annals of Statistics}, 40(2):1263--1282.

\bibitem[Nagelkerke et~al., 2000]{nagelkerke2000estimating}
Nagelkerke, N., Fidler, V., Bernsen, R., and Borgdorff, M. (2000).
\newblock Estimating treatment effects in randomized clinical trials in the
  presence of non-compliance.
\newblock {\em Statistics in Medicine}, 19(14):1849--1864.

\bibitem[Permutt, 1990]{permutt1990testing}
Permutt, T. (1990).
\newblock Testing for imbalance of covariates in controlled experiments.
\newblock {\em Statistics in medicine}, 9(12):1455--1462.

\bibitem[Romano and Wolf, 2016]{romano2016efficient}
Romano, J.~P. and Wolf, M. (2016).
\newblock Efficient computation of adjusted p-values for resampling-based
  stepdown multiple testing.
\newblock {\em Statistics \& Probability Letters}, 113:38--40.

\bibitem[Russell and G., 1995]{Davidson1995estimation}
Russell, D. and G., M.~J. (1995).
\newblock Estimation and inference in econometrics.
\newblock {\em Econometric Theory}, 11(3):631--635.

\bibitem[Seaman and White, 2013]{seaman2013review}
Seaman, S.~R. and White, I.~R. (2013).
\newblock Review of inverse probability weighting for dealing with missing
  data.
\newblock {\em Statistical methods in medical research}, 22(3):278--295.

\bibitem[Sichieri and Cunha, 2014]{sichieri2014unbalanced}
Sichieri, R. and Cunha, D.~B. (2014).
\newblock Unbalanced baseline in school-based interventions to prevent obesity:
  adjustment can lead to bias-a systematic review.
\newblock {\em Obesity facts}, 7(4):221--232.

\bibitem[Solmi et~al., 2014]{solmi2014permutation}
Solmi, F., Onghena, P., Salmaso, L., and Bult{\'e}, I. (2014).
\newblock A permutation solution to test for treatment effects in alternation
  design single-case experiments.
\newblock {\em Communications in Statistics-Simulation and Computation},
  43(5):1094--1111.

\bibitem[Tang et~al., 2009]{tang2009applying}
Tang, L., Duan, N., Klap, R., Asarnow, J.~R., and Belin, T.~R. (2009).
\newblock Applying permutation tests with adjustment for covariates and
  attrition weights to randomized trials of health-services interventions.
\newblock {\em Statistics in medicine}, 28(1):65--74.

\bibitem[Tanizaki, 2004]{tanizaki2004small}
Tanizaki, H. (2004).
\newblock On small sample properties of permutation tests: Independence between
  two samples.
\newblock {\em International Journal of Pure and Applied Mathematics.},
  13:235--244.

\bibitem[Terenzini, 1982]{terenzini1982designing}
Terenzini, P.~T. (1982).
\newblock Designing attrition studies.
\newblock {\em New directions for institutional research}, 1982(36):55--71.

\bibitem[Wei and Zhang, 2001]{wei2001analysis}
Wei, L. and Zhang, J. (2001).
\newblock Analysis of data with imbalance in the baseline outcome variable for
  randomized clinical trials.
\newblock {\em Drug information journal}, 35(4):1201--1214.

\bibitem[Welch, 1990]{welch1990construction}
Welch, W.~J. (1990).
\newblock Construction of permutation tests.
\newblock {\em Journal of the American Statistical Association},
  85(411):693--698.

\bibitem[Winkler et~al., 2014]{winkler2014permutation}
Winkler, A.~M., Ridgway, G.~R., Webster, M.~A., Smith, S.~M., and Nichols,
  T.~E. (2014).
\newblock Permutation inference for the general linear model.
\newblock {\em Neuroimage}, 92:381--397.

\bibitem[Wooldridge, 2010]{wooldridge2010econometric}
Wooldridge, J.~M. (2010).
\newblock {\em Econometric analysis of cross section and panel data}.
\newblock MIT press.

\end{thebibliography}

\end{document}